\documentclass[referee]{aa} 
\usepackage{graphicx}
\usepackage{txfonts}
\begin{document}
   \title{Analysis of the carbon-rich very metal-poor dwarf
      G77-61
\thanks{based in part on observations obtained at the W.M. Keck Observatory, 
which is operated jointly by the California Institute of Technology, 
the University of California, 
and the National Aeronautics and Space Administration.}}

   \subtitle{ }

   \titlerunning{The carbon-rich metal-poor dwarf G77-61}
   \authorrunning{Plez \& Cohen}
   
   \author{Bertrand Plez
          \inst{1}
          \and
          Judith G. Cohen\inst{2}}

   \offprints{B. Plez}

   \institute{GRAAL, CNRS UMR~5024, Universit\'e Montpellier II, F-34095
                 Montpellier cedex 5, France
              \email{plez@graal.univ-montp2.fr}
         \and
            Palomar Observatory, Mail Stop 105-24, 
            California Institute of Technology, Pasadena, Ca. 91125
             }

   \date{Received ; accepted }

   \abstract{
We present an analysis of the carbon dwarf G77-61,
a known binary,
based on high resolution Keck spectra.
G77-61 has a very-low metallicity, although not as extreme
as what was previously conjectured.
This star is very carbon-enhanced, the spectra showing very strong
CH, CN, and C$_2$ bands of both $^{13}$C and $^{12}$C isotopes.
Atomic lines are sparse, and often blended, but we 
were able to  derive abundances
for Na, Mg, Ca, Cr, Mn, and Fe.
Its [Fe/H] of -4.03$\pm$0.1 places it among the lowest metallicity
stars known, and its very high [C/Fe]=+2.6  , and [N/Fe]=+2.6  
among the most C and N-rich metal-poor stars. The carbon isotopic ratio is
$^{12}$C/$^{13}$C=5$\pm$1. No overabundance of s-
or r-process elements is detectable, but the upper limits that can be set 
on these overabundances are not very constraining. 
   }

   \maketitle
%

\section{Introduction \label{section_intro} }
The star G77-61 was first identified as a nearby carbon dwarf by 
Dahn et al. (\cite{dahn}).  They reanalyzed its previous
parallax determination by Routly (\cite{routly}) to find $M_V = +9.6\pm0.6$ mag.
Dearborn et al (\cite{dearborn}) established that G77-61 is a binary with an unseen
companion and with a period of $\sim$245 days, suggesting a cool white dwarf
of much higher mass than the visible object as the second star. 
The primary was later spectroscopically analysed by Gass et al. (\cite{gass}), who 
concluded that its metallicity was extremely low, [Fe/H]=$-$5.5 dex (on our adopted scale
with logA(Fe)=7.45).
No other spectroscopic 
investigation has been carried out since then. 

The discovery of a large number
of very metal-poor, carbon-rich objects, with diverse additional peculiarities, 
particularly
$r$-process or/and $s$-process enrichment, and the discovery of the
most iron-poor star known, \object{HE0107$-$5240}
(Christlieb et al. \cite{christlieb}), at [Fe/H]=$-$5.3, which is also very 
C-rich, all contribute
to a renewed interest in the star G77-61.

\section{Observations}
The star G77-61 was observed several times using HIRES (\cite{vogt94})
at the Keck Observatory.
The first set of spectra (3 exposures, each 1200 sec long)
were obtained on the night of Sep 29, 2002.  They covered a spectral
range from 3780 to 5320~\AA, at a resolution of 34\,000
 with small gaps between the orders
at the red end of the range.  A 1.1 arcsec wide slit was used,
and a signal to noise ratio (SNR)
 of 100/spectral resolution element (4 pixels wide) in the
continuum in the middle of this spectral range was achieved.
This was calculated assuming Poisson statistics and ignoring
issues of cosmic ray removal, flattening etc.
The star was re-observed August 19, 2003 to cover the region from
5000 to 7900~\AA, with a resolution of 34\,000.
Three separate exposures were required for full
spectral coverage.
On line signal calculations used the order just blueward of the
center of the CCD at a wavelength of 5865 A.
Much shorter exposure times were required to achieve
the desired SNR of 75 to 100/spectral resolution element
as the star is so red.

These spectra were reduced using a combination 
of Figaro scripts (\cite{shortridge93}) and
the software package MAKEE\footnote{MAKEE was developed
by T.A. Barlow specifically for reduction of Keck HIRES data.  It is
freely available on the world wide web at the
Keck Observatory home page, http://www2.keck.hawaii.edu:3636/.}.
Normal procedures were followed until the final step, which involves
fitting a low order polynomial to the continuum to remove the
last residuals of the variation of the instrumental response with
wavelength.  Since the continuum is essentially impossible
to determine in this heavily banded star, even at this spectral resolution,
this step was omitted.  Typical residuals removed at this stage are
about 10\% of the mean continuum within an order and are locally
smooth.

The observed 
heliocentric radial velocity for G77-61 is $-29$ km/s 
on September 29, 2002, and $-20$~km/s
on August 19, 2003, with somewhat large uncertainty due to the strong veiling.
These values fit very well with those predicted from the orbit of Dearborn et al.
(\cite{dearborn}): -29.6 and -18.5~km/s.

\section{Analysis}
The spectra were analysed through spectrum synthesis, the very strong veiling
by molecular bands not allowing equivalent width determinations.

\subsection{stellar parameters}
Our B-V photometry from the Swope 1-m telescope at the
Las Campanas Observatory, (Ivans \& Cohen, private communication),
 B=15.70 and V=13.97, gives B$-$V=1.73 mag.
Dahn et al. (\cite{dahn}) provide V=13.90, B$-$V=1.70. The trigonometric parallax
of 16.9$\pm$2.2mas (Harris et al. \cite{harris}), and the V magnitude
lead to M$_{\rm V}$=10.1$\pm$0.27, assuming no reddening, as the star is nearby.
Schlegel et al.'s extinction maps (\cite{schlegel98}) give E(B-V)=0.100 at this position.
The Galactic latitude of G77$-$61 is $-41.51^o$, corresponding to a height of 
39~pc above the Galactic plane, resulting in about 25\% of the total extinction, or 0.025~mag,
well below all other uncertainties.
The IR colours J$-$H=0.684, H$-$K=0.337, and J$-$K=1.007
in the Johnson system (Bessell \& Brett \cite{bessell}) are obtained
from 2MASS photometry (Skrutskie et al. \cite{2mass1}, and Cutri et al. \cite{2mass2}) 
J=11.470 (0.022), H=10.844 (0.024), K=10.480 (0.019),
using the relations of Carpenter (\cite{carpenter}).
 The K magnitude is 10.52 in the Johnson system, which transforms
to an absolute M$_{\rm K}$=6.62$\pm$0.27. Carbon-rich dwarf MARCS models around 
$T_{\rm eff}$=4000~K
and [Fe/H]=$-3$ or $-4$ have bolometric corrections BC$_{\rm K}\approx$2.25,
and BC$_{\rm V}\approx-0.65$. Thus, the absolute bolometric magnitude is 
M$_{\rm bol}\approx$9.45 from V, and 8.87 from K. Dearborn et al. (\cite{dearborn})
find BC$_{\rm V}=-0.96$ from spectro-photometric observations, 
leading to M$_{\rm bol}$=9.14. We adopt 
the average of the K and V values, M$_{\rm bol}$=9.16
(i.e. L=0.017~L$_{\odot}$).

We adopt $T_{\rm eff}=4000~\pm200~K$, obtained from interpolating the observed V-J and V-K
colors in the predicted color grid of \cite{houdashelt00}.
We assume that the star is sufficiently nearby that the reddening
is negligible.  
The uncertainty in T$_{\rm eff}$
is not due to the observational photometric errors but rather to the modeling
uncertainties.  The \cite{houdashelt00} color grid was calculated
assuming scaled Solar compositions extending to [Fe/H] as low as $-3$ dex,
but G77--61 has a highly anomalous composition with very large C excess.
The C$_2$ bands reduce the flux integrated over the $V$ bandpass below
that of the true continuum by $\sim$35\%.  The effect of the prominent
C$_2$ absorption, which is not included
in the models at all, has been compensated for in an approximate way,
but still leaves a large uncertainty in $T_{\rm eff}$.  The range of
$T_{\rm eff}$ from the three previous analyses of this star listed in \S\ref{section_intro}
is 4000 to 4200~K.

For $T_{\rm eff}$=4000~K  we deduce that
R=0.27~R$_{\odot}$, and $\log~g$=5.27 assuming M=0.5M$_{\odot}$,; 
assuming M=0.3M$_{\odot}$ leads to
$\log~g$=5.05.

While evolutionary tracks for metal poor stars
with such  extreme C enhancement are not available,
 the new Yale-Yonsei
isochrones (Kim et al. \cite{kim}, Yi et al. \cite{yi})
with $\alpha$-enhancement give an indication
of whether these values are in accord with those
expected for an old lower main sequence star.  We examine
the sequence of metal poor isochrones with $\alpha$-enhancements
of a factor of four and age of 9Gyr.  The inferred luminosity of G77-61
then corresponds to a mass of 0.38M$_{\odot}$.  The $T_{\rm eff}$ of the
isochrone at this luminosity, which is probably quite uncertain,
is $\sim$300~K lower than the value we adopt.

\subsection{model atmospheres and line data}

Using the estimates of $T_{\rm eff}$, and $\log~g$, models were calculated 
with the MARCS code (Gustafsson et al. \cite{gust75}, Plez et al. \cite{pbn92},
Edvardsson et al. \cite{edvardsson},
Asplund et al. \cite{asplundmarcs}, Gustafsson et al. \cite{gust03}), 
at metallicities between $-5$ and $-3$, 
for various C/H ratios. The synthetic spectra were computed 
with the TURBOSPECTRUM package (Alvarez and Plez \cite{alvaplez}), 
using atomic lines from
the VALD database (Kupka et al. \cite{vald}). The data for the lines we use in the abundance
analysis is listed by Hill et al. (\cite{hill}), and \cite{cayrel}.
The molecular line lists for CH, and CN were
specially assembled.  
These lists, used by Hill et al. (\cite{hill}), were further updated. 
The CH line list was improved
by shifting a number of line positions by  0.1 \AA\  or less between
4200 and 4275\AA , for $^{13}$CH A-X lines, and 
to a lesser extent $^{12}$CH lines
(using in particular Richter \& Tonner, \cite{rich}, as well
as a Keck spectrum of the carbon-rich metal-poor star HE0212-0557).
The C$_2$ line list was taken from Querci et al. (\cite{querci}), and updated 
by scaling the gf-values to agree with modern lifetime and intensity
measurements (Cooper and Nicholls \cite{cooper}, Grevesse et al. \cite{grevesse91},
Erman and Iwamae \cite{erman},
Langhoff et al. \cite{langhoff}). The situation is far from satisfactory for spectrum
synthesis with this latter list, and efforts are under way to improve
the situation. We did not use C$_2$ lines to derive the C abundance, and avoided 
the use of other lines situated within C$_2$ bands, where a risk of blending exists.

\begin{figure}
   \centering
   \resizebox{\hsize}{!}{\includegraphics{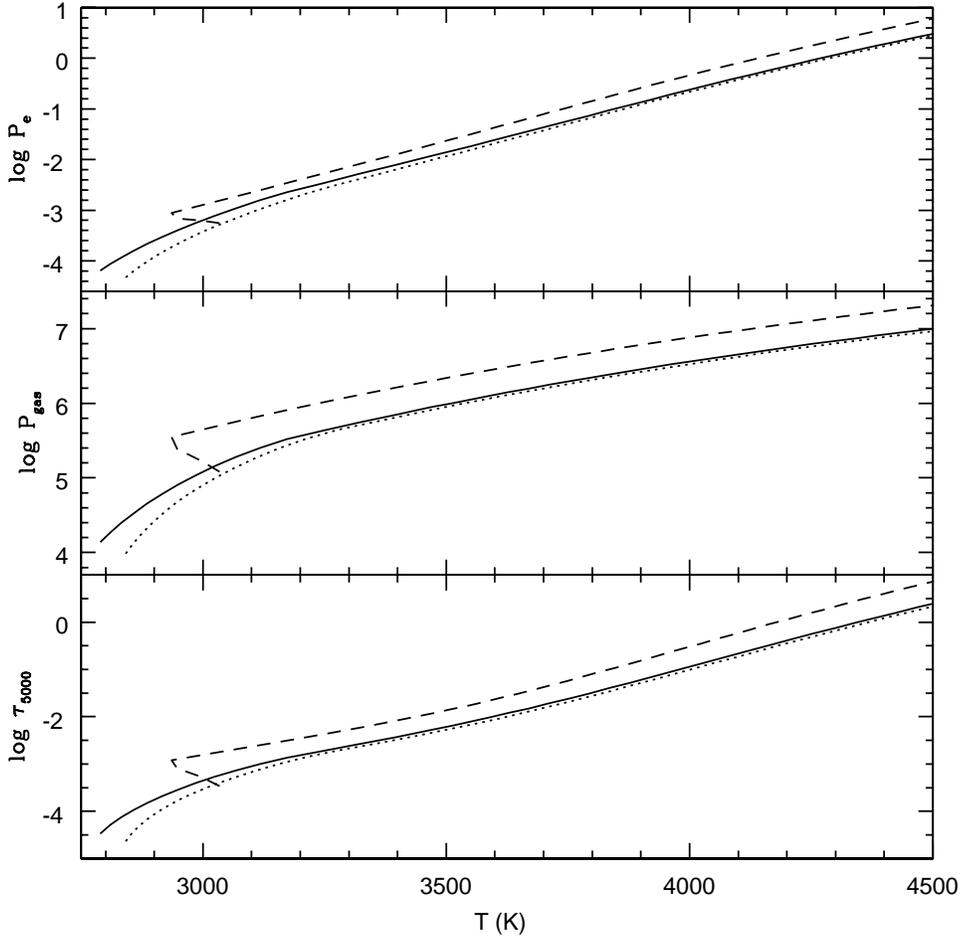}}
   \caption{Effect of abundance changes on model atmosphere
   thermal and pressure structure. Full line is the final model, with the abundances
   of Table \ref{Table_abund},  dashed line is a model with solar abundances scaled
   by $-4$~dex, and 
   [$\alpha$/Fe]=0.4, dotted line has in addition CNO abundances to their final
   value.}
   
             \label{FigureG77_tp}
  \end{figure}

\begin{figure}
   \centering
   \resizebox{\hsize}{!}{\includegraphics{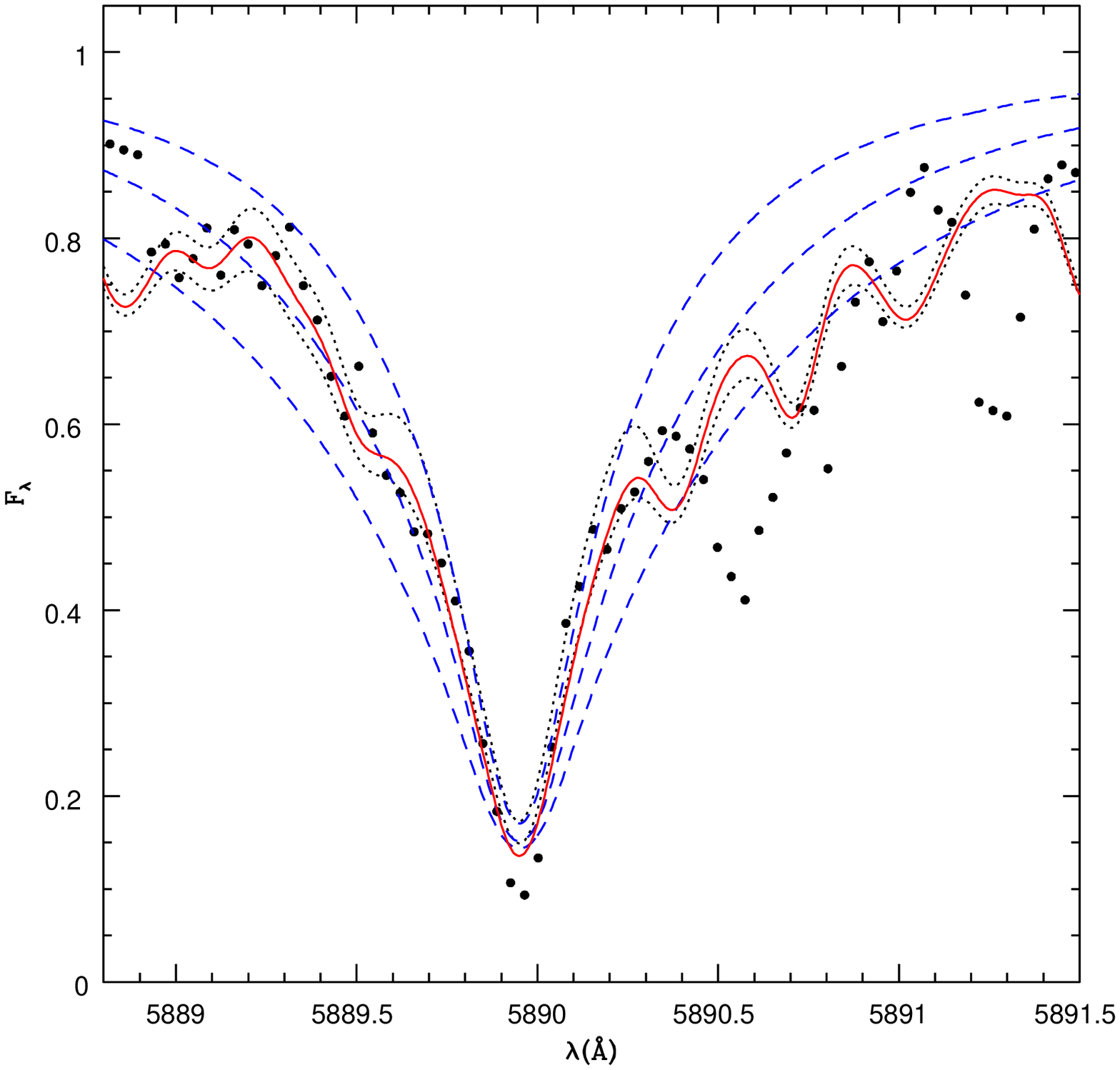}}
   \caption{Effect of model atmosphere changes on the Na~I D line.
  Dots are observations; full line is the synthetic spectrum 
   from the model atmosphere with the final abundances, including 
   logA(Na)=2.92; dotted lines are for 2 Na abundances: 
   2.92 and 3.22, and a model with solar abundances scaled
   by $-4$~dex, except CNO which have their final value, and 
   [$\alpha$/Fe]=0.4 (dotted line model of Fig.~\ref{FigureG77_tp});
   finally, dashed lines are for Na abundances of 2.32, 2.62 and 2.92
   and the model with solar abundances scaled by $-4$~dex 
   (dashed line model of Fig.~\ref{FigureG77_tp}). The features in the observed spectrum that are not matched by our calculations are due to C$_2$.}
             \label{FigureG77_NaI}
  \end{figure}

It is important to use a model computed with abundances close to the 
spectroscopic determination. 
Figure \ref{FigureG77_tp} shows the thermal and pressure structure 
of the final model adopted, with abundances from Table \ref{Table_abund},
together with a second model with solar abundances uniformly scaled 
down by $-4$~dex and with [$\alpha$/Fe]=+0.4,
and a third model  with in addition CNO abundances set to their final values.
The thermal structure of the model is affected by the abundance of species 
that contribute importantly to the opacity (mostly C and N, due to the predominance of C$_2$ and CN in the spectrum), and less importantly
 by the abundance of 
electron donors (Mg, Na, and Ca) which has a impact on the T-P$_e$-P$_{gas}$
 relation in the outermost layers.
This is further illustrated in the spectrum synthesis.
Changing from a model with scaled
solar abundances ([X/H]=$-4.0$), and [$\alpha$/Fe]=$+0.4$,
 to a model with the abundances of Table \ref{Table_abund}
results for example in a change of more than 0.3 dex
in the determination of the abundance of Na
from the Na~I D line. This is illustrated in Figure \ref{FigureG77_NaI} 
where synthetic spectra computed
with various Na abundances in the three model atmospheres are compared 
to the observation. In the case of the two models with equal 
abundances of C, N, and only differences in some elements (esp. Na, Mg, Ca),
the differences are due to the impact of the electron donors 
in the line forming layers. They resulting difference is about 0.1~dex in the 
determination of the Na abundance.
Interestingly, deeper in the atmosphere, around $\tau_{\rm Ross}$=1,
about 80\% of the electrons arise from H$_3^+$. The obvious impact this has on H$^-$
opacity demonstrates the importance of including H$^+_3$ in the molecular equilibrium,
as recently shown for the evolution of zero-metal stars by Harris et al. (\cite{harris04}).
We use the partition function of Neale \& Tennyson (\cite{neale}) for H$^+_3$. 

The effect of changes in C and N  abundances on the emergent 
spectrum is even more dramatic,
through the change in blanketing of C$_2$, CN and CH, which affects the
thermal structure of the model (see Figure \ref{FigureG77_tp},
and \ref{FigureG77_NaI}).
 The O abundance,
which we cannot determine, has no impact as long as it stays well
below the C abundance. Low-resolution IR spectra (Joyce \cite{joyce})
as well as IR photometry (Dahn et al. \cite{dahn}) show very faint
CO bands, in favor of a low oxygen abundance.
We adopted here logA(O)=5.0.

\subsection{results}

\begin{figure}
   \centering
   \resizebox{\hsize}{!}{\includegraphics{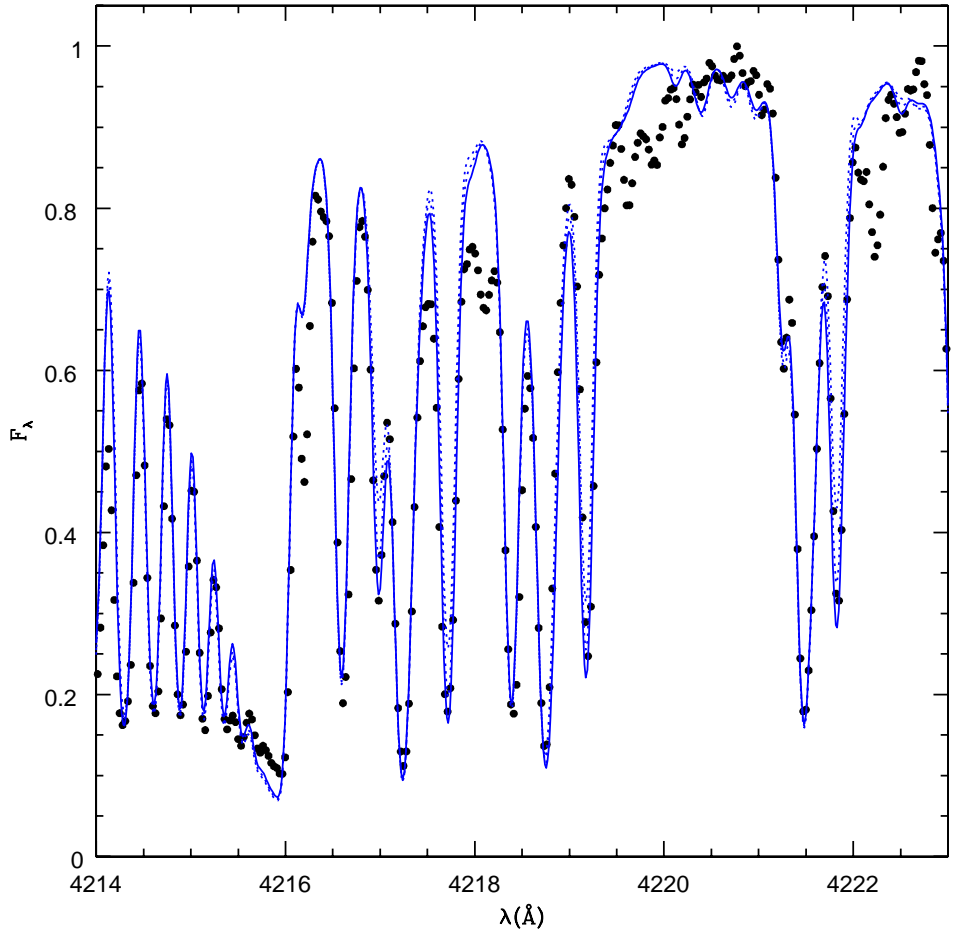}}
    \label{FigureG77_CH}
   \caption{CH A-X band head. Dots are the observed spectrum.
    Synthetic spectra
   were calculated for $^{12}$C/$^{13}$C=5 (final adopted value), 9, and 20.
   All abundances have their final value.}
  \end{figure}

\begin{figure}
   \centering
   \resizebox{\hsize}{!}{\includegraphics{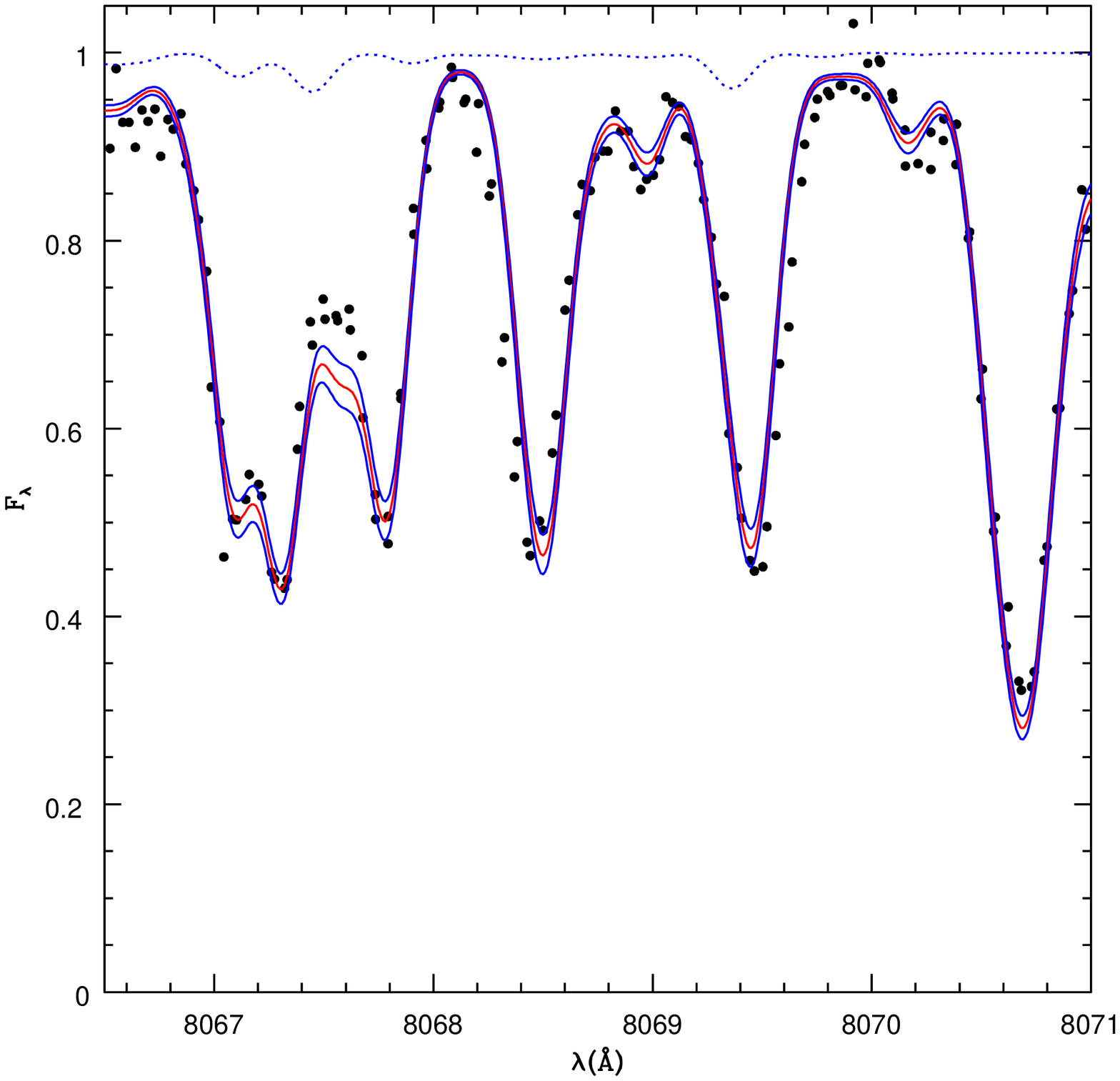}}
   \caption{CN red system lines. Dots are two observed spectra, with different setups.
    Synthetic spectra
   were calculated for 3 different N abundances, logA(N)=6.32, 6.42, 6.52, and
   without CN lines.}
             \label{FigureG77_CN}
  \end{figure}

\begin{figure}
   \centering
   \resizebox{\hsize}{!}{\includegraphics{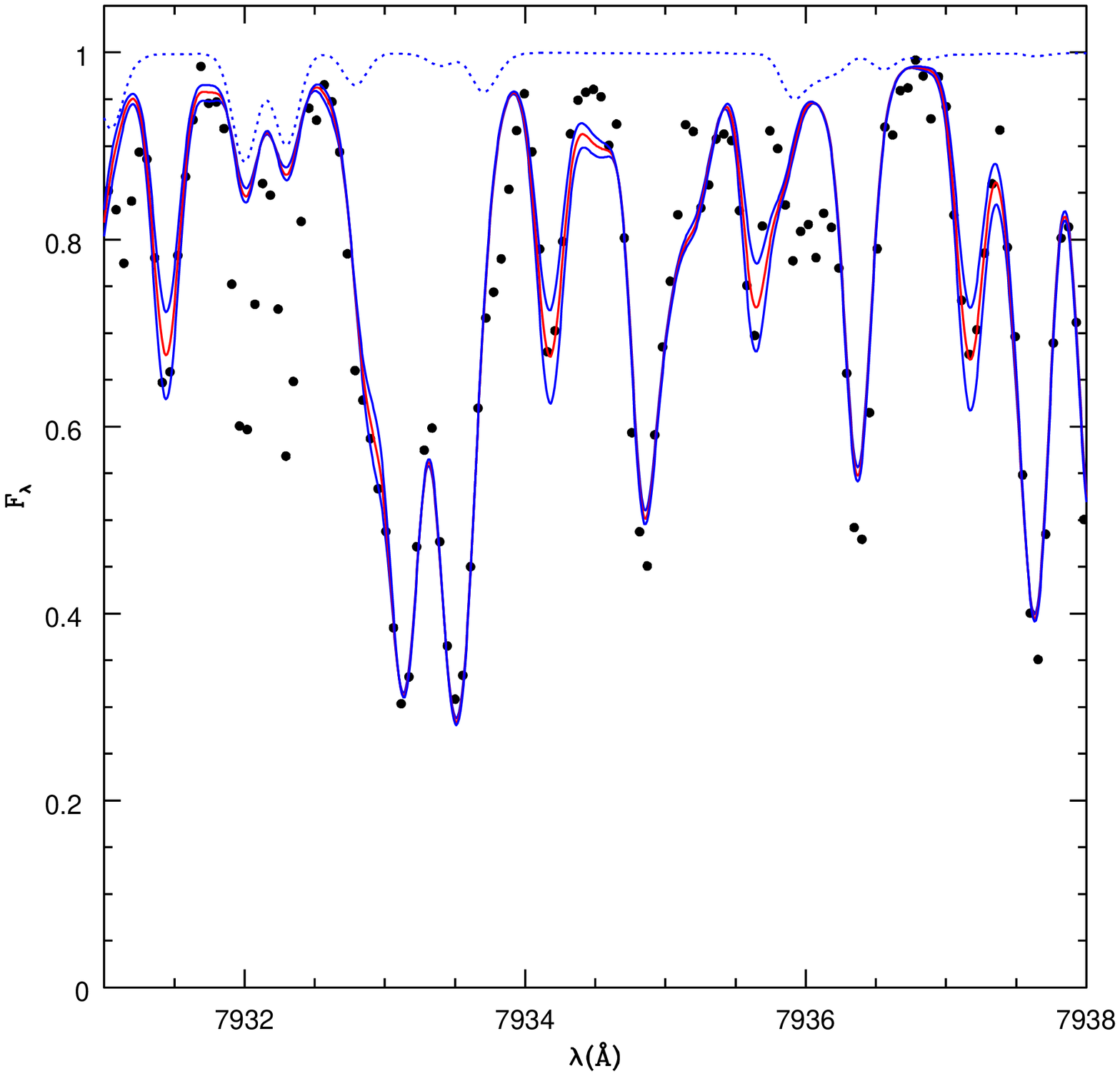}}
   \caption{Red system lines of $^{12}$CN and $^{13}$CN. Spectra were computed
   for 3 different isotopic ratios: $^{12}$C/$^{13}$C=4, 6, 9, as well as without 
   any CN lines.
   The (too) faint spectral lines appearing in the latter case (dotted line)
   are C$_2$ lines from the Phillips transition system. The line list we use
   for C$_2$ is not reliable.}\label{FigureG77_1213CN}
  \end{figure}

Abundances were first derived for C using
the CH A-X and B-X bands around  3890 and 4100-4200\AA\  (see Figure \ref{FigureG77_CH}),
 with checks on the C$_2$
Swan bandheads at 5160\AA\  and 4735\AA. 
 The N abundance was 
estimated using the CN A-X red bands around 8000\AA. Note that 
an increase of the derived carbon abundance would lead to a decrease of
the nitrogen abundance by the same amount.
The C$_2$ Swan and Phillips bands, although very strong and displaying
obvious bandheads for the $^{12}$C$_2$, the $^{12}$C$^{13}$C, and
$^{13}$C$_2$ isotopic combinations, could not be used, because of 
the absence of reliable line list.
The $^{12}$C/$^{13}$C isotopic ratio  of 5$\pm$1, was derived using 18 $^{13}$CN
lines between 7925 and 7966\AA. Wavelengths of $^{13}$CN lines from
the list of Plez were shifted by 0.1\AA\ to improve the match to the observations.
Figure \ref{FigureG77_CN}
and Figure \ref{FigureG77_1213CN} show parts of these regions.
The synthetic spectrum without CN lines included is also plotted. Faint lines 
appear in this synthetic spectrum, which seem well placed but too faint to fit 
the observed spectrum. These lines are low excitation C$_2$ lines from the Phillips
bands, for which we don't have a reliable line list. Efforts to improve the situation
are ongoing, as these unsaturated lines seem promising to determine the 
carbon abundance in stars similar to G77-61.

The impact of microturbulence is not negligible on the 
strong saturated lines.We could only use a handful of strong atomic lines. 
Most molecular lines are also somewhat saturated.
We could thus  not derive a microturbulence parameter from the observations.
We decided to use 0.1km/s for the following reason:
in the Sun the ratio of microturbulence parameter to the maximum MLT velocity 
is 1.3/2.5. The MLT velocity in our C dwarf model is 0.22km/s, reached 
around $\tau_{5000}$=1. Assuming the same ratio as for the Sun, 
which is reasonable if
we suppose that microturbulence measures convective overshooting 
in the upper layers of the atmosphere, 
 we find a microturbulent velocity
of the order of 100m/s, which is one order of magnitude smaller than
the thermal velocity of atoms at 3000K. 
Actually, convection reaches optically thin layers in the carbon dwarf models
with velocities of about 100m/s at  $\tau_{5000}$=0.01, which 
reinforce our estimate for the microturbulence parameter.
Using 1km/s would not impact the 
abundances derived here  very much. Only the use of a very unlikely
microturbulence parameter
of the order of 2km/s or more would induce significant abundance changes,
and a notably worse match to the observations.

Unblended atomic lines are rare in the spectrum of G77-61, and most are 
strong.
The abundances we present are 
based on few lines, often with low excitation energies, and subject
to non-LTE effects. Large errors might have occurred due to an inappropriate 
description of the upper layers of the stellar atmosphere.
We avoided lines in regions of the spectrum where the synthetic spectra
did not reproduce well enough the molecular veiling. In particular,
 the region below 4000\AA\ shows numerous large absorption features,
of unidentified origin. These features show up in the spectra of other  cool C-rich,
metal-poor stars as well, and efforts toward their identification should
be made. 

The abundances are presented in Table \ref{Table_abund}, where we use as 
reference solar abundances the values from Grevesse and Sauval 
(\cite{grevesse}), except for C=8.41, N=7.80, and O=8.67
 which we take from \cite{allende02}, \cite{asplund03}, 
 and \cite{asplund04}, and Fe=7.45 from \cite{asplund00}.
 Our solar abundances are the same as Christlieb at al. (\cite{christlieb04})
  and \cite{cohen04}.

\begin{table}
\caption{Abundances derived for G77-61\label{Table_abund}}
\begin{tabular}{llllll}
Element&$\log\epsilon$(X)&error&[X/H]&[X/Fe]&wavelength (\AA)\\
\hline  
C (CH)$^*$  &7.0  &$\pm$0.1  &-1.4  &+2.6   &\\
N (CN)      &6.4  &$\pm$0.1  &-1.4  &+2.6   &\\
Li~I        &$<$1 &          &$<$0  &$<$4   &6708\\
Na~I        &2.9  &$\pm$0.2  &-3.4  &+0.60  &5889\\
Mg~I        &4.0  &$\pm$0.2  &-3.5  &+0.49  &5183, 5172\\
K~I$^b$     &2.3  &$\pm$0.3  &-2.8  &+1.21  &7665, 7699\\
Ca~I        &2.7  &$\pm$0.2  &-3.7  &+0.37  &4226\\
Ca~II       &2.9  &$\pm$0.2  &-3.5  &+0.57  &8498, 8542, 8662\\
Cr~I        &2.0  &$\pm$0.2  &-3.7  &+0.36  &5206, 5208\\
Mn~I        &1.3  &$\pm$0.3  &-4.1  &-0.06  &4030\\
Fe~I        &3.42 &$\pm$0.15 &-4.03 &       &3887, 3895, 3899\\
            &     &          &      &       &4045, 4063, 4071\\
Zn~I        &     &          &      &$<$2.7 &4810\\
Sr~II       &     &          &      &$<$0   &4077\\
Ba~II       &     &          &      &$<$1   &4554, 4934\\
Eu~II       &     &          &      &$<$3?  &4129, 4205\\
\hline
\multicolumn{6}{l}{$^a$ $^{12}$C/$^{13}$C=5$\pm$1}\\
\multicolumn{6}{l}{$^b$ The K abundance is quite uncertain: blend with CN}
\end{tabular}
\end{table}

\begin{figure}
   \centering
   \resizebox{\hsize}{!}{\includegraphics{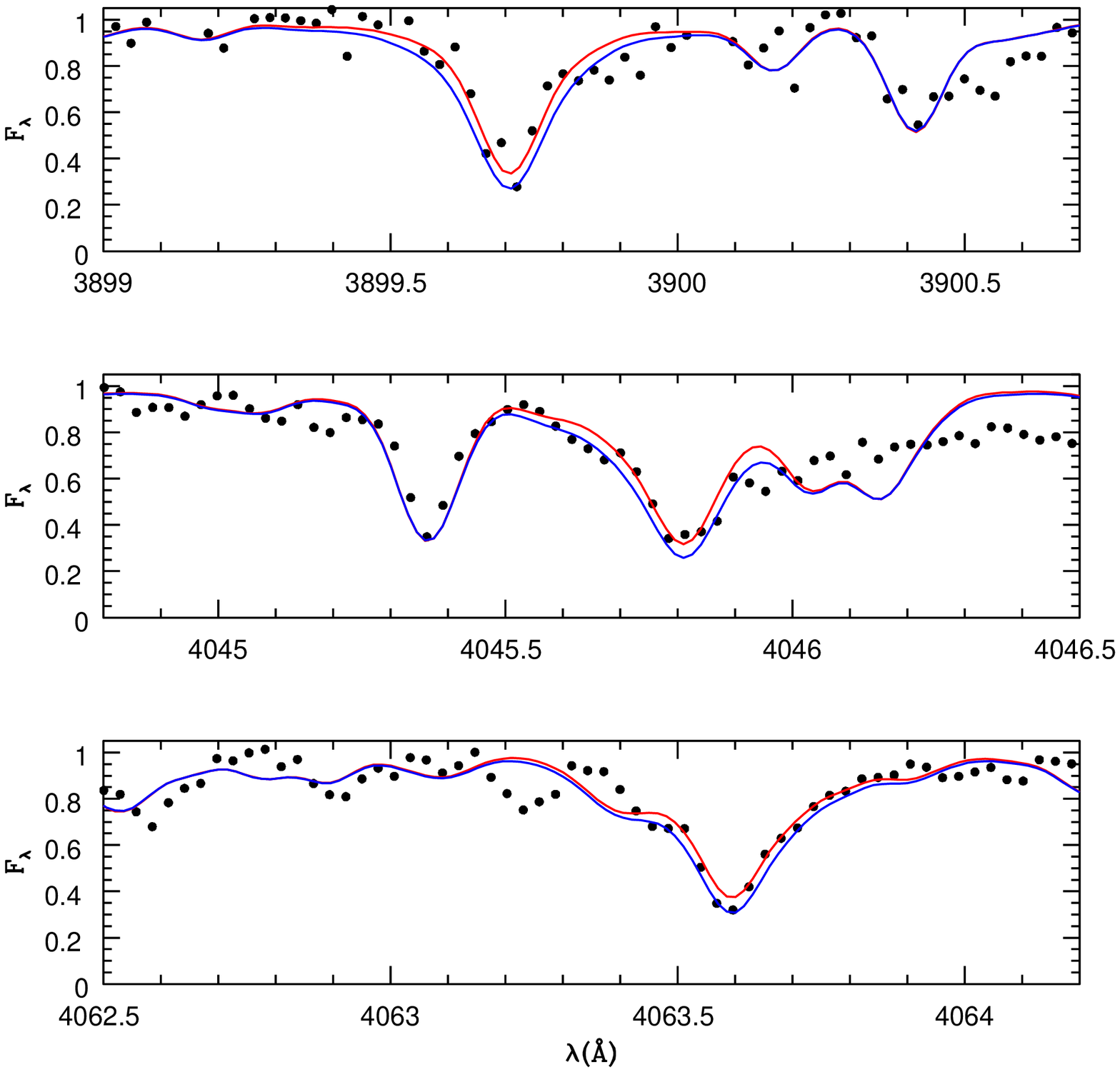}}
   \caption{Observed (dots) and calculated spectra (full lines) of 3 Fe~I lines, for
   2 Fe abundances:
   3.30 and 3.60, using the final model atmosphere parameters.}\label{FigureG77_FeI}
  \end{figure}

The uncertainties from the quality of the fit of individual lines  are of
the order of 0.15~dex. In some cases (K~I) possible additional 
blends increase the uncertainty
to greater values (0.3 dex or more). For Fe~I 
(see Fig. \ref{FigureG77_FeI}), the situation is 
better, as 6 lines could be used, that are not resonance lines, 
contrary to Na, or K. The standard deviation from the 6 lines is 0.1~dex.
The uncertainties quoted in Table \ref{Table_abund} are estimated on 
this basis. They do not include any systematic error 
that could arise from a change of stellar parameters. 
The abundances derived
using the model with solar abundances scaled down by 4~dex, except the alpha elements
([$\alpha$/Fe]=+0.4), and the final CNO abundances, differed from the abundances
quoted in Table \ref{Table_abund} by 0~dex for Ca, by less than 0.1~dex for Fe, 
by +0.15~dex for Mg and Cr, by  0.3~dex for Mn, and 0.6~dex for Na~I. At the upper limit to the $T_{\rm eff}$ for G77-61, i.e. 4200~K, [Fe/H] becomes -3.85.  This
is a reasonably secure upper limit to the metallicity of this star.
An increase of the gravity to $\log g = 5.5$ reduces the Fe abundance
by 0.3~dex. For Ca~II, the changes are $-0.1$~dex when the gravity is increased
to  $\log g = 5.5$, and $+0.3$~dex for an increase of T$_{\rm eff}$ to 
4200~K, while the Na~I abundance rises by $+0.5$~dex.
These values 
give an estimate of the systematic errors that may affect our results, not accounting for possible non-LTE effects.

\section{Discussion}

We have shown that G77-61 is an extremely metal poor dwarf with [Fe/H] $-4.03$~dex
which is highly C-enhanced, having $N$(C)/$N$(Fe)  500 times the Solar ratio,
so that [C/H] is $-1.4$~dex.    There are only $\sim$10 stars known
in the Galaxy with
[Fe/H $< -3.5$~dex verified by a detailed abundance 
analysis\footnote{These are
\object{CD$-$38$^{\circ}$245}, \object{CS~22162--002}, \object{CS~22885--096}
 and \object{CS~22949--037},
studied in detail by \cite{norris01}, \object{CS~22172--002}, \object{CS~22968--014}
from the sample of \cite{cayrel}, \object{CS~29498--043} \cite{aoki02a}, 
\cite{aoki02b},\cite{aoki04},
\object{BS~16545--0089} and \object{HE~0218--2738} from
the sample of \cite{cohen04},
and \object{HE~0107--5240}, the most Fe-poor star known \cite{christlieb}.}.
\object{G77--61} is arguably the third most Fe-poor star known in the Galaxy today; only
\object{HE~0107--5240} is significantly more metal poor.  It is interesting that many
of these stars are also extremely C-enhanced
\footnote{\object{CD$-$38$^{\circ}$245} is the most metal poor star 
 known which is not extremely C-enhanced.};  
\object{HE~0107--5240} would
appear as a C-star, as does \object{G77--61}, were it a cool dwarf rather than
a subgiant with a much hotter $T_{\rm eff}$.  Furthermore, \object{G77--61} is the only
star in this group which was originally noticed through its high proper motion;
all the others were found in spectroscopic surveys.

Results from large recent surveys of very metal poor stars (Cayrel et al, 2004 for giants,
and Cohen et al, 2004 for dwarfs) have now been
published.  \cite{cohen04} have pointed out that 
the abundance ratios [X/Fe] for elements between Mg and Zn, while they
may be functions of [Fe/H], are constant to within the observational error
at a fixed value of [Fe/H].  They also provide evidence that this
is true of most C-rich stars, ignoring the CNO elements of course.
Because of the complex spectrum of G77-61 with its many strong molecular
bands including C$_2$, we could only determine abundances for a limited
set of elements.  These are compared with the mean abundances derived
for very metal poor giants and for dwarfs with normal CNO
from \cite{cayrel} and \cite{cohen04} in  Table~\ref{table_comp}.
\begin{table}
\caption{Comparison of G77-61 with Normal Extremely Metal Poor Stars\label{table_comp}}
\begin{tabular}{lllll}

          &Sun   &{G77-61}   &mean dwarfs$^a$  &mean giants$^b$  \\
Element&$\log\epsilon$(X)    &[X/Fe] &[X/Fe]    &[X/Fe]\\
\hline                                          
C (CH)    &8.39   &+2.6      &           &           \\
N (CN)    &7.80   &+2.6      &           &          \\
Li~I      &1.10   &$<$4      &           &            \\
Na~I      &6.33   &+0.60     &           &-0.24     \\  
Mg~I      &7.54   &+0.49     &+0.52      &+0.23     \\ 
K~I       &5.12   &+1.21     &           &+0.27   \\   
Ca~I      &6.36   &+0.37     &+0.31      &+0.22    \\   
Ca~II     &6.36   &+0.57     &           &    \\                                          
Cr~I      &5.67   &+0.36     &-0.21      &-0.51         \\
Mn~I      &5.39   &-0.06     &-0.55      &-0.52$^c$  \\
Fe~I      &7.45   &-4.03$^d$ &-2.7$^e$   &-3.01$^e$\\
Zn~I      &4.60   &$<$2.7    &+0.52      &+0.48    \\     
Sr~II     &2.97   &$<$0      &-0.44$^f$  &     \\
Ba~II     &2.13   &$<$1      &-0.32      &   \\
Eu~II     &0.51   &$<$3      &           &   \\
\hline
\multicolumn{5}{l}{$^a$ Sample of Cohen et al. 2004}\\
\multicolumn{5}{l}{$^b$ Sample of Cayrel et al. 2004, from their regression at [Fe/H]=-4.0}\\
\multicolumn{5}{l}{$^c$ The NLTE 0.4~dex correction Cayrel et al. uses has been removed.}\\
\multicolumn{5}{l}{$^d$ [Fe/H]}\\
\multicolumn{5}{l}{$^e$ median [Fe/H] of the sample} \\
\multicolumn{5}{l}{$^f$ corrected for difference of reference solar Sr}\\
\end{tabular}
\end{table}
To within the errors, G77--61 appears to also show similar abundance ratios,
apart from the lighter elements discussed further below.  
There are however discrepancies: 
 [Mn/Fe] is $\sim$0.5 dex lower in the normal
giants and dwarfs than it is in G77--61. [Cr/Fe] is more than 0.5~dex
lower in the dwarfs, and 0.8~dex in the giants. The case of [K/Fe] almost
1~dex higher in G77--61 than in the mean giants, may be due to the difficulty
of measuring the resonance K~I lines affected by telluric absorption, 
and strongly blended with molecular lines. Despite this, the overall 
agreement between G77-61 and the mean metal-poor giants and dwarfs
is remarkably good considering the difficulty of the analysis.

The abundance ratios in C-rich metal poor stars, which obey the above relations 
for the elements between Mg and the Fe-peak, appear to float for the lighter
and for the heavier elements beyond the Fe-Peak.
Some extremely metal poor C-stars have large enhancements of the heavy elements, i.e.
Ba, La and even Pb in some cases
(see, e.g., \cite{vaneck}, \cite{aoki02c}, \cite{lucatello03},
\cite{sivarani04} and \cite{cohen04}).
Those dwarfs with $s$-process enhancement
must arise in binary systems with a more massive primary, now
presumably a white dwarf, in which mass transfer has occured.  
Other very metal poor
C-rich stars show no enhancement of the heavy elements
(see, e.g. Aoki et al. 2002c or Cohen et al. 2004).
Although we only have rough upper limits to the abundance of elements heavier
than the Fe peak in G77--61, it appears
to be one of the latter group, with no or only modest enhancement of the
heavy elements. Regarding mass-transfer, Wallerstein \& Knapp 
(\cite{wallerstein}) conjectured that the extremely low metallicity of G77--61
could be due to a loss of non-volatile elements during the mass transfer.
This hypothesis does not seem supported by any observation. The 
metallicity now
determined around 1/10000 solar is not as extreme as previously 
believed, and the general pattern of abundances is close to other dwarfs
and giants of similar metallicity. In addition, at such a high C/O ratio
the dust formed would be graphite or amorphous carbon, not incorporating 
Mg, Ca, Fe, and other elements that condensate in e.g. silicates in
O-rich environments.   

With regard to the CNO elements, the general trend 
seen by the second author in examining the spectra of large numbers of
very metal poor stars is that
if C is highly enhanced, then N may be enhanced, but the range of enhancement
varies significantly.  On the other hand, N is never highly enhanced unless
there is a substantial C-enhancement.  We compare the deduced abundances
for G77--61 with those of four very metal poor C-enhanced-stars 
in Table~\ref{table_compC}, 
the first of which is HE~0107--5240, the most Fe-poor star known, analyzed
by Christlieb et al. (\cite{christlieb04}).
There are two independent detailed abundance
analyses of \object{CS22949--037}, a highly C-enhanced, extremely 
metal-poor giant,  by \cite{norris01}
and by Depagne et al. (\cite{depagne}).  The agreement between them indicates the level of
accuracy one can expect for stars with such complex spectra.  However
\object{G77--61} is much cooler than this star, and its spectrum is 
correspondingly more complex.
\object{CS~22957--027} is another extremely metal poor giant with a 
very large C-enhancement studied by \cite{norris97} and by \cite{aoki02a};
the results of the more recent analysis are shown in the table.
\object{CS~29498--043} (\cite{aoki02a}, \cite{aoki02b}, \cite{aoki04})
is a very metal poor giant with highly enhanced Na and Mg
as well as CNO.
\object{HE~0007-1832} is a very metal poor dwarf with no
enhancement of the heavy elements (\cite{cohen04}). 

\begin{table*}
\caption{Comparison of G77--61 with carbon-rich Extremely Metal Poor Stars\label{table_compC}}
\begin{tabular}{lllllllll}
            &G77--61   &CS29498--043 &HE0107--5240 &\multicolumn{2}{c}{CS22949--037} &CS22957--027 &HE0007--1832 \\
\hline
            &[X/Fe]        &[X/Fe]$^a$ &[X/Fe]     &[X/Fe]   &[X/Fe]$^a$   &[X/Fe]$^a$     &[X/Fe]\\                
\hline
Li~I        &$<$+4         &           &$<$+5.3    &         &$<$-0.3  &           &       \\
C (CH)      &+2.6          &+2.20$^b$  &+3.70      &+1.28    &+1.27    &+2.4       &+2.55\\
$^{12}$C/$^{13}$C &5       &6$^c$      &$>$50      &         &4        &8          &7     \\
N (CN)      &+2.6          &+2.26      &+2.70$^a$  &+2.84    &+2.66    &+1.62      &+1.85\\
O (CO)      &              &+2.45      &+2.4$^d$   &         &+2.12    &           &        \\
Na~I        &+0.60         &+1.41      &+0.81      &         &+2.09    &           &         \\
Mg~I        &+0.49         &+1.74      &+0.15      &+1.21    &+1.35    &+0.65      &+0.76\\
Ca~I        &+0.37         &+0.10      &-0.09      &+0.40    &+0.31    &+0.14      &+0.32\\
Ca~II       &+0.57         &           &+0.36      &         &         &           &        \\              
Cr~I        &+0.36         &-0.41      &$<$+0.26   &-0.60    &-0.44    &           &        \\
Mn~I        &-0.06         &-1.00      &$<$+0.36   &-0.97    &-0.84    &-0.41      &-0.23\\
Fe~I$^e$    &-4.03         &-3.49      &-5.28      &-3.74    &-3.94    &-3.12      &-2.65  \\   
Zn~I        &$<$2.7        &$<$+0.5    &$<$+2.65   &         &+0.63    &           &        \\
Sr~II       &$<$0          &-0.57      &$<$-0.52   &-0.02    &+0.25    &-0.56      &+0.07\\
Ba~II       &$<$1          &-0.42      &$<$+0.82   &-0.89    &-0.61    &-1.23      &+0.16  \\     
Eu~II       &$<$+3?        &           &$<$+2.78   &$<$+0.88 &$<$0.01  &$<$+0.83   &$<$+1.79  \\
\hline
Source      &              &f          &g          &h        &i        &c          &j\\
\hline
\multicolumn{8}{l}{$^a$ Corrected for differences in solar reference abundance}\\
\multicolumn{8}{l}{$^b$ from C$_2$}\\
\multicolumn{8}{l}{$^c$ Aoki et al. 2002a}\\
\multicolumn{8}{l}{$^d$ Bessell, Christlieb \& Gustafsson, 2004}\\
\multicolumn{8}{l}{$^e$ [Fe/H]}\\
\multicolumn{8}{l}{$^f$ Aoki et al. 2004}\\
\multicolumn{8}{l}{$^g$ Christlieb et al. 2004}\\
\multicolumn{8}{l}{$^h$ Norris, Ryan \& Beers, 2001}\\ 
\multicolumn{8}{l}{$^i$ Depagne et al. 2002}\\
\multicolumn{8}{l}{$^j$ Cohen et al. 2004}\\
\end{tabular}
\end{table*}

Table~\ref{table_compC} gives an indication of the 
variety found among the C-rich metal poor stars.
The four stars listed have $N$(C)/$N$(N) ranging over more than a factor of 1000, with
no obvious correlation with [Fe/H]. 
G77--61 and all of the comparison stars, except \object{HE0107-5240},
though presumably a giant (Christlieb et al. \cite{christlieb04}),
show low C$^{12}$/C$^{13}$ ratios, indicating substantial CN processing.
If one ignores the CNO elements, a crucial
distinction among them involves the elements just heavier than CNO, the easiest of
which to detect are Na and Mg.    If those elements are not enhanced, then
the abnormality is confined to the enhancement of the CNO elements, and an origin
involving normal processes of nucleosynthesis within stars of intermediate mass
can be imagined; for example a binary scenario
with no heavy element production is viable.  However, if Na and Mg are enhanced, as is the
case especially for CS~22949--037 and for CS~29498--043, and to a lesser extent
 for the other three stars, then conventional wisdom suggests that
significant production  of Mg and Na is not expected for intermediate mass metal-poor
stars, and a supernova event is required.
It is these arguments that have led to the discussion of SN with fallback models
for extremely metal poor massive stars \cite{umeda03} and \cite{tsujimoto}
where most of the lighter elements, produced in the outer zones, 
get ejected, but the mass
cut is such that the heavier elements fall back onto the collapsed core.  The abundance ratios of the ejecta can be fine-tuned 
by invoking mixing as well as fallback.

However, very recent caclulations by \cite{herwig05} and by \cite{karakas05}
demonstrate that intermediate mass very low metallicity AGB stars
can produce and mix to their surface substantial amounts of Na and Mg, as well
as  copious amounts of C and N, and a low $^{12}$C/$^{13}$C ratio due to
CN processing. This eliminates
the need to seek a Type II SN origin for their excesses in G77--61 and in the
other stars listed in Table~\ref{table_compC}.  
In addition, according to \cite{goriely2004}, the s-process efficiency
may be very much reduced  in low metallicity AGB stars, due to the higher
temperature of the bottom of the convective zone. This could explain the 
class of very metal-poor carbon-rich stars without excess of heavy elements
to which \object{G77--61} seems to belong.

We may thus view the O abundance as the key to separating excesses of the light
elements C, N, Na, Mg as arising from Type II SN of some kind versus binary
mass transfer involving an extremely low metallicity AGB star.  In the former
case, copious O should be present, while in the latter, O should not be
significantly enhanced.  \object{HE0107--5240} falls into the former category,
as do at least two of the comparison stars.  But preliminary information
regarding the O abundance in G77--61 
based on the low-resolution IR spectra of Joyce (\cite{joyce}) is that the O abundance is
low. However, Joyce ascribed the apparent weak CO 
both to the low metallicity of G77--61 and 
at least partially to suppression of near-IR molecular bands by 
collision-induced absorption (H$_2$-H$_2$, and to a lesser extend He-H$_2$).
Our model atmosphere calculations with high O abundance, and including 
collision-induced absorption predict strong CO bands. 
In any case all SN models tend to produce large amounts of O, relative to C,
and it is therefore crucial to determine the O abundance in \object{G77 -- 61}
to separate the binary AGB from the SN hypothesis for the 
selective abundance enhancements in this extremely metal-poor dwarf carbon
star.
A high resolution near-IR spectrum is already in hand and results should
be available in the near future.

\begin{acknowledgements}
We thank N. Christlieb for suggesting a re-investigation of this star.
B. Freytag is thanked for suggesting the microturbulence estimate, 
and R. Cayrel and E. josselin for useful comments.
The entire Keck/HIRES user communities owes a huge debt to 
Jerry Nelson, Gerry Smith, Steve Vogt, and many other 
people who have worked to make the Keck Telescope and HIRES  
a reality and to operate and maintain the Keck Observatory. 
We are grateful to the W. M.  Keck Foundation for the vision to fund
the construction of the W. M. Keck Observatory. 
The authors wish to extend special thanks to those of Hawaiian ancestry
on whose sacred mountain we are privileged to be guests. 
Without their generous hospitality, none of the observations presented
herein would have been possible.
JGC is grateful to the National Science Foundation for partial support under
grant AST-0205951.
This publication makes use of data from the Two Micron All-Sky Survey,
which is a joint project of the University of Massachusetts and the 
Infrared Processing and Analysis Center, funded by the 
National Aeronautics and Space Administration and the
National Science Foundation.
\end{acknowledgements}

\end{document}